\documentclass{ws-ijmpcs}

\begin{document}

\markboth{Raymond R. Volkas}
{Pangenesis: a common origin for visible and dark matter}

%
\catchline{}{}{}{}{}
%

\title{PANGENESIS: A COMMON ORIGIN FOR VISIBLE AND DARK MATTER
}

\author{RAYMOND R. VOLKAS}

\address{ARC Centre of Excellence for Particle Physics at the Terascale, School of Physics\\
The University of Melbourne, Victoria, 3010, Australia\\
raymondv@unimelb.edu.au}

\maketitle

\begin{history}
\received{Day Month Year}
\revised{Day Month Year}
\end{history}

\begin{abstract}
The similar mass densities observed for visible and dark matter in the present-day universe suggest a common origin for both.  A scheme called ``pangenesis'' for realising this using the Affleck-Dine mechanism in a baryon-symmetric universe is presented in this talk. 

\end{abstract}

\ccode{PACS numbers: 98.80.Cq, 95.35.+d, 12.60.Jv, 11.30.Fs, 11.30.Qc}

\section{Introduction}	

The universe today is composed of three main components.  By mass-density, ordinary or visible matter (VM) accounts 
for about 4\%, dark matter (DM) for 21\% and dark energy for the remainder.  The purpose of this talk is to suggest an 
explanation for why the densities of visible and dark matter are so similar, with the latter about a factor of five more
dense than the former.   If dark matter is supposed to be an independent sector from ordinary matter, then their rather
similar mass densities is a puzzle.  The generic expectation is that the mass densities should differ by orders
of magnitude.  But we may turn this around and explore the hypothesis that the similar densities imply a deep connection
between VM and DM, especially the way both are produced in the early universe.  In this talk, I report on a scheme 
developed with Bell, Petraki and Shoemaker for producing visible and dark matter from a common source, and with 
tightly related number densities.\cite{pangenesis}

The starting point is the fact that the ordinary matter density today is due to a baryon-antibaryon asymmetry.  If the
plasma of the early universe had been exactly symmetric between particles and antiparticles, then the very efficient 
annihilation processes would have left a universe today with hardly any matter in it.  
It is thus necessary to suppose that the primordial plasma had a small asymmetry in favour of particles
over antiparticles, with the matter in the universe today comprised of the excess baryons and electrons.

If we wish to explain why the DM density today is similar to VM, then it makes sense to explore the idea that the present
DM population is also the relic from a slightly asymmetric primordial dark plasma.  This general idea has been
discussed for a long time, but has recently attracted increased attention as well as having acquired the catchy title of 
`asymmetric DM' (ADM).\cite{See}  
It should be contrasted with the much-studied hypothesis that DM is a Majorana fermion,
such as the neutralino or the sterile neutrino.  The relic abundance of a Majorana particle obviously cannot be due to
a particle-antiparticle asymmetry (unless the mass is tiny so that the two helicity states become effectively particle and
antiparticle).

Making DM asymmetric renders it qualitatively similar to ordinary matter, but we want to go further 
and make a quantitative
connection.  Because a mass density depends on both a number density and a particle mass, we will ultimately need
a way to relate these two properties between the two sectors.  In this talk, I discuss how one may relate number
densities, leaving the particle mass question open.  The key idea is that the universe may be secretly baryon symmetric!
It could be that there is a well-defined sense in which DM carries antibaryon number which exactly cancels the baryon
number carried by the VM.\cite{BS}  
If true, then the baryon asymmetry is an illusion.  Nevertheless, there is a genuine asymmetry
in another quantum number, as we now discuss. (Actually, there is a certain poetic licence inherent in the 
``baryon-symmetric universe'' title.  The precise definition appears below.)

\section{The baryon-symmetric universe}

The stability of ordinary matter is due in significant measure to the conservation of baryon number.  For our purposes,
it is convenient to rename this quantity as `visible baryon number' and designate it as $B_1$.  It is this quantity we refer
to when talking of the baryon asymmetry of the universe.  In ADM models, there is an analogous `dark baryon number'
$B_2$ that plays an important role in ensuring the stability of DM, and its relic density is due to a $B_2$ asymmetry.

Now form the orthogonal linear combinations
\begin{equation}
B \equiv B_1 - B_2,\qquad X \equiv B_1 + B_2,
\end{equation}
where $B$ is a generalised baryon number carried by both VM and DM.  For simplicity, this will simply be called `baryon
number'. The baryon-symmetric universe refers to this quantity, postulating that $\Delta B = 0$ for the universe as a
whole, with
\begin{equation}
\Delta B_1 = \Delta B_2 \neq 0.
\end{equation}
The orthogonal charge $X$ is necessarily asymmetric:
\begin{equation}
\Delta X = \Delta B_1 + \Delta B_2 = 2 \Delta B_{1,2} \neq 0.
\end{equation}
The point of this is that DM number density is determined by $\Delta B_2$ and the baryon number carried by the DM
particle or particles.  Since $\Delta B_2$ equals the visible baryon asymmetry, we achieve a strong connection between
the number densities of VM and DM, which is our goal.

We now know what we want, but how do we realise it?  The simple equations above tell us the answer:  we need to
generate an $X$ asymmetry while maintaining $\Delta B=0$.  Rather than generating visible baryon number 
\emph{per se}, we want to generate $X$ number.  The Sakharov conditions then become our guide.\cite{Sakharov}  
We want
early universe dynamics that features $X$-, $C$- and $CP$-violating interactions that occur out of equilibrium.  Since
we do not wish a $B$ asymmetry, then $B$ must either be always conserved, or violated without the other conditions
holding.  Elegance and robustness argues for the $B$ conservation option: the universe is baryon symmetric, because
$B$ is not violated by the dynamics that creates the particle asymmetries.  The $X$-creation process must switch off
in the late universe, so that $B_1$ and $B_2$ are independently conserved today.

\section{Pangenesis}

One may now take any standard mechanism for baryon asymmetry generation, and recast it for $X$ generation.
In this talk, I describe how one may use the Affleck-Dine (AD) mechanism\cite{AD}
in a scenario termed ``pangenesis'', meaning the unified creation of all matter\cite{pangenesis} 
(see also Ref.~[\refcite{CZ}]).

In a nutshell, the AD mechanism is the coherent production of charge (equivalent to an asymmetry) from the oscillations
of a scalar condensate when the associated symmetry is explicitly broken.  Recall that the Noether current for a
charged scalar field is $J^\mu = i ( \phi^* \partial^\mu \phi - \partial^\mu \phi^* \phi)$.  The charge density is $J^0$,
which can be expressed as
\begin{equation}
J^0 = \rho^2 \dot{\theta},
\end{equation}
where $\phi \equiv (\rho/\sqrt{2}) \exp(i \theta)$.  The creation of a charge-carrying scalar condensate requires it to be
oscillating in time.  The initiation of these oscillations requires the Sakharov conditions to be met.

Because $X$-violation today must be extremely weak, we need a way of amplifying these effects in the early universe.
The AD pangenesis structure is this:  At the renormalisable level, both $B$ and $X$ are conserved.  Effective
non-renormalisable operators that preserve $B$ but violate $X$ are then introduced.  These effective operators
are suppressed by some powers of an ultraviolet cutoff scale $M$, with the exponents depending on the specific 
operators used.  The key observation is that the suppression will be lifted if scalar fields appearing in those
operators develop large values in the early universe.  

For a scalar field to be able to acquire a large value, it must define a
flat direction in the scalar potential.  Flat directions are a generic feature of scalar potentials in 
supersymmetric (susy) theories,
so scenarios invoking the AD mechanism fit in very well with the susy solution to the gauge hierarchy problem.  The
exactly flat directions found in generic renormalisable theories with unbroken supersymmetry are lifted by both
soft susy-breaking scalar masses and non-renormalisable effective operators.  Nevertheless, the directions remain
flat enough for the AD mechanism to work: the soft electroweak-scale masses are negligible in the high-energy regime
where AD condensate formation occurs, and the non-renormalisable terms only come into play for exceptionally large
field values.

Of the many possible models for pangenesis, I shall describe only one.\cite{pangenesis}  
This model has three sectors: visible, dark
and connector.  The visible sector is an extension of the minimal susy standard model (MSSM).  The dark sector
is described by a separate gauge theory and will not be specified in this talk; there are many possibilities.  The connector
sector is where the AD dynamics takes place.  It contains flat-direction scalar fields that form an $X$-laden condensate.
Through $B$ and $X$ conserving interactions, the $X$ charge of the condensate is transferred into a $B_1$-asymmetric
visible particle plasma and a $B_2$-asymmetric dark-sector soup.

Consider Table \ref{ta1}.  It lists the chiral superfields $\Phi_j = (\phi_j, \psi_j, F_j)$,
and there are also vector-like partners $\hat{\Phi}_j = (\hat{\phi}_j, \hat{\psi}_j, \hat{F}_j)$.  
The U(1) charges are arranged so that $\Phi_0$ and $\hat{\Phi}_0$ carry $X$-charge but zero $B$-charge.  These two
superfields define the connector sector.  The supermultiplets $\Phi_1$ and $\Phi_2$ (and their vector-like partners) 
belong to the visible and dark sectors, respectively.  Their role is to transfer the $X$ asymmetry carried by the
connector fields to the other two sectors.

\begin{table}[t]
\tbl{Visible and dark baryon number assignments}
{\begin{tabular}{@{}cccc@{}} \toprule
\ & $\Phi_0$ & $\Phi_1$ & $\Phi_2$ \\
 \colrule
$B_1$ & -1 & 1 & 0 \\
$B_2$ & -1 & 0 & 1 \\
$B$ & 0 & 1 & -1 \\
$X$ & -2 & 1 & 1\\ \botrule
\end{tabular} \label{ta1}}
\end{table}

The renormalisable superpotential for the superfields in Table \ref{ta1} is
\begin{equation}
\delta W_r = \kappa \Phi_0 \Phi_1 \Phi_2 + \hat{\kappa} \hat{\Phi}_0 \hat{\Phi}_1 \hat{\Phi}_2
+ \sum_j \mu_j \Phi_j \hat{\Phi}_j.
\end{equation}
We see therefore that there are no quartic terms for the $\phi_0, \hat{\phi}_0$ scalar potential, so these
fields define a ``flat plane'', a two-dimensional generalisation of a flat direction.  There are of course potentially
many other flat directions, especially when one includes the MSSM and dark sector fields.  One then needs a reason
for why large field values are generated in the $\phi_0, \hat{\phi}_0$ plane, but not in other directions.  An elegant
solution to this problem lies in having the role of $B_1$ played by ordinary baryon-minus-lepton number -- call it
$(B-L)_1$ -- rather than just ordinary baryon number.  Because $(B-L)_1$ is anomaly-free, we can gauge it.  Let us
suppose that actually the generalised ``baryon'' number $B = (B-L)_1 - B_2$ is gauged, but $X$ is not.  Now, to
get a flat direction, we need both $D$-flatness and $F$-flatness.  With gauged $B$, $D$-flatness requires that
flat directions carry a vanishing expectation value for the $B$ charge.  This lifts all potential flat directions that could
spoil pangenesis, because we are now assured that any flat direction is incapable of generating an asymmetry in $B$.

To have the baryon-symmetric AD mechanism operate, we now add non-renormalisable terms to the superpotential 
which respect $B$ but explicitly violate $X$.  Examples include $\Phi_0^4$, $\Phi_0^3 \hat{\Phi}_0$,
$\Phi_0 \hat{\Phi}_0^3$ and $\hat{\Phi}_0^4$.  There are others that preserve both $B$ and $X$, but they do
not play any important role in the AD dynamics.  These quartic superpotential terms are multiplied by $\lambda/M$, where
$\lambda$ is a dimensionless coupling constant and $M$ is the UV cut-off as before.  The coupling constants contain
the required $CP$-violating phases, as well as helping to quantify the strength of explicit $X$ violation.

The important terms in the scalar potential for the flat plane are\cite{pangenesis}
\begin{eqnarray}
V_{\rm AD} 
&=&\! \left[m_0^2(T) - c H^2\right] |\phi_0|^2  + \left[\hat{m}_0^2(T)  -  \hat{c} H^2\right] |\hat{\phi}_0|^2  \nonumber \\ 
&+& \sum_{k=0}^4 \frac{(A_k \tilde{m} + a_k H) \lambda_k}{M} \phi_0^k \hat{\phi}_0^{4-k} \nonumber \\
&+& \sum_{k=0}^3 \sum_{l=0}^{3-k} \frac{\lambda^2_{kl}}{M^2} 
\left(\phi_0^* \hat{\phi}_0\right)^{3-k-l}|\phi_0|^{2k}|\hat{\phi}_0|^{2l} + h.c., 
\label{eq:VFD}
\end{eqnarray}
where  $m_0^2(T) \simeq \tilde{m}^2 + \kappa^2 T^2$ include thermal masses, $H$ is the Hubble parameter, and soft
susy breaking is parameterised by $\tilde{m} A_k$.  The Hubble parameter enters the potential because the
finite vacuum energy density during inflation breaks susy.  This potential has all the ingredients for AD generation
of $X$, since the terms in the second and third lines violate both $X$ and $CP$.  The vacuum-energy induced
$c H^2$ corrections to the scalar masses are also very important, because they allow the initial scalar field values
after inflation to be very large. The coefficients $c$ and $\hat{c}$ are required to be positive, so that when $H^2$
gives the dominant contribution to the squared masses the potential minima are displaced to very large field values.
During inflation, the scalar fields fall into these minima.  The subsequent evolution has two aspects to it because we
have a two-field AD setup, rather than the standard one-field situation.  
The evolution can be decomposed into the angular and radial directions.  Being $X$ and $CP$ violating, 
the sextic terms in the third line of 
Eq.\ (\ref{eq:VFD}) are able to drive $X$ creation through evolution along the angular direction, before the standard AD
radial oscillations are able to begin.  This is useful, because the thermal masses are known to have a suppressing
effect on asymmetry generation for the case where the explicit charge violation occurs at quartic order in the
superpotential.  The radial evolution begins after $H^2$ decreases sufficiently to make
the squared masses positive.  The angularly-evolving fields now no longer have to maintain a large radius because
that is no longer energetically favoured.

Explicit calculations\cite{Twofield} have shown that the asymmetry generated is given by
\begin{equation}
\eta(X) \sim 10^{-10} \left( \frac{\sin\delta}{\lambda} \right) \left( \frac{T_R}{10^9\ {\rm GeV}} \right) \frac{M}{M_P},
\end{equation}
where $\eta$ is the number density asymmetry relative to entropy density, 
$\delta$ is a measure of the $CP$-violating phases, $T_R$ is the reheating temperature after inflation
and $M_P$ is the Planck mass.  The correct asymmetry can be obtained for reasonable choices of all parameters.

The $X$ asymmetry held by the condensate must now be transferred into both visible and dark sector particles.
This occurs through the $\Phi_0 \Phi_1 \Phi_2$ and $\hat{\Phi}_0 \hat{\Phi}_1 \hat{\Phi}_2$ renormalisable
terms. Since these terms respect both $B$ and $X$, their role is purely one of transfer; they cannot change the value
of the asymmetry.  Recall that $\Phi_1$ and its partner are visible sector fields, and $\Phi_2$ and its partner
reside in the dark sector.  The scalar fields $\phi_0$ and $\hat{\phi}_0$ defining the condensate can decay directly
into the fermionic components of $\Phi_{1,2}$ and $\hat{\Phi}_{1,2}$.  We see that for every visible sector field
created, there is a corresponding dark sector field, so the key baryon-symmetric feature is preserved.  The 
supermultiplet $\Phi_1$ may couple in many ways to MSSM fields and thus inject the asymmetry throughout the
visible sector: $\Phi_1 L H_u$, $\Phi_1 L L \bar{e}$, $\Phi_1 L Q \bar{d}$ and $\Phi_1 \bar{u} \bar{d} \bar{d}$.
Similarly $\Phi_2$ and/or $\hat{\Phi}_2$ may couple to further dark-sector degrees of freedom.  The asymmetry
in the visible sector will in general be reprocessed by sphaleron effects in the usual way.  For example, if the
$\Phi_1 L H_u$ term is the means by which the transfer occurs, then the $X$-charge initially manifests as
lepton number, and the sphaleron reprocessing is essential.

For the dark asymmetry to be the sole determinant of the DM number density, we need the symmetric part of the dark
plasma to efficiently annihilate into radiation that subsequently redshifts and becomes insignificant.  This means
a dark sector force of some description must exist.  A simple possibility is that it is a dark analogue of electromagnetism:
an unbroken U(1) gauge force.  This has the added benefit that the dark photon could kinetically mix with the
ordinary photon and enhance the direct detection prospects of the asymmetric DM.  The U(1) gauge symmetry can
also be spontaneously broken.

The cosmological observations of the visible and dark matter densities imply a prediction for the dark matter mass.
It is given by
\begin{equation}
m_{\rm DM} = q_{\rm DM} \frac{\Omega_{\rm DM}}{\Omega_{\rm VM}} \frac{\eta(B_1)}{\eta(B_2)} m_p,
\end{equation}
where $m_p$ is the proton mass, and $q_{\rm DM}$ is the baryon number of the DM particle.  For reasonable values
of the latter parameter, we see that the several-GeV regime is predicted.  This is common to all ADM models, with
the baryon-symmetric variety providing an explicit value for $\eta(B_1)/\eta(B_2)$.  Note that this is typically
not equal to one, because of sphaleron reprocessing.  The analogue of sphaleron effects may well exist in the
dark sector as well, so that needs to be specified before the ratio of the final asymmetries can actually be computed.
Conversely, a terrestrial measurement of the DM mass would tell us what $\eta(B_2)$ has to be and thus help us
construct the internal physics of the dark sector.  It is interesting that present results from DAMA\cite{DAMA}, 
CoGeNT\cite{CoGeNT} and CRESST\cite{CRESST}
favour a DM mass in this range, though of course the true situation is far from settled.  Finally, we note that
the above discussion needs to be generalised if the DM is multicomponent, which is a perfectly reasonable possibility.

This scenario has potential implications for both DM direct detection experiments and collider physics.  If $B$ is
gauged, then the associated $Z'$ mediates spin-independent DM-nucleon scattering, with a per nucleon 
cross-section given by
\begin{equation}
\sigma^{\rm SI}_{B-L} \simeq (4 \times 10^{-44}\, {\rm cm}^2)\, q^2_{\rm DM} \left( \frac{g_{B-L}}{0.1} \right)^4 
\left( \frac{0.7\ {\rm TeV}}{M_{B-L}} \right)^4.
\end{equation}
For the case where a massive dark photon $D$ kinetically mixes with the ordinary photon, the cross-section is
\begin{equation}
\sigma^{\rm SI}_{D} \simeq (10^{-40}\,  {\rm cm}^2) \left( \frac{\epsilon}{10^{-4}} \right)^2 \left( \frac{g_{D}}{0.1} \right)^2 
\left( \frac{1\ {\rm GeV}}{M_{D}} \right)^4.
\end{equation}
This is potentially large enough to be relevant for the DAMA and CoGeNT signals, but there is enough parameter freedom
to lower it should that become necessary (to obey the XENON constraints\cite{XENON}, for example).  
The $Z'$ could be directly produced at the LHC if it is light enough.  If so,
it would have a substantial invisible width into dark sector species, making it distinguishable from standard $Z'$ 
candidates.\cite{Qua}

Finally, we comment that the lightest supersymmetric particle in the visible sector should contribute in a negligible way
to the DM density in pangenesis.  If the visible LSP is a stable neutralino, for example, it should be primarily an 
admixture of the wino and the higgsinos to allow it to annihilate efficiently and be underabundant.

\section{Conclusion}

Pangenesis -- the simultaneous creation of visible and dark matter in a baryon-symmetric universe via the Affleck-Dine
mechanism -- is a viable cosmological scenario that is consistent with the supersymmetric resolution of the gauge
hierarchy problem.  Evidence for pangenesis would be: a DM mass in the several GeV range, supersymmetry,
a $Z'$ with significant coupling to the dark sector, and a DM direct detection process mediated by kinetic mixing
between the ordinary photon and a dark-sector analogue.

\section*{Acknowledgments}

This work was supported in part by the Australian Research Council.  I thank K. Petraki for useful comments on an
early draft of this proceedings article.

\end{document}